\documentclass[11pt]{article}
\usepackage[T1]{fontenc}
\usepackage[latin1]{inputenc}
\usepackage{graphicx}
\usepackage{bm}
\usepackage{amsmath}
\usepackage{lastpage}
\usepackage{fancyhdr}
\usepackage{color}
\usepackage{pdfpages}
\usepackage{picins}

\pdfoutput=1

\begin{document}

\title{\textbf{Fluctuation Relations for Diffusions Thermally Driven\\
by a Non-Stationary Bath}}
\date{}
\author{Raphaël Chetrite \\
%EndAName
{\small Physics of Complex Systems, Weizmann Institute of Science,} \\
{\small Rehovot 76100, Israel}\\
{\small and} \\
{\small Laboratoire de Physique, C.N.R.S., ENS-Lyon, Universit\'e de Lyon,} 
\\
{\small All\'ee d'Italie, 69364 Lyon, France}}
\maketitle

%\author{Rapha$\mathrm{\ddot{e}}$l Chetrite \\
%EndAName
%\\
%Laboratoire de Physique, C.N.R.S., ENS-Lyon, Universit\'e de Lyon, \\
%46 All\'ee d'Italie, 69364 Lyon, France}

\abstract{\noindent In the context of the dynamical evolution in a non-stationary 
thermal bath, we construct a family of fluctuation relations for the entropy 
production that are not verified by the work performed on the system.
We exhibit fluctuation relations which are global versions either of 
the generalized Fluctuation-Dissipation Theorem around a non-equilibrium 
diffusion or of the usual Fluctuation-Dissipation Theorem for energy 
resulting from a pulse of temperature.}

\section{Introduction}

One important recent progress in non-equilibrium statistical physics is the
discovery of various fluctuation relations which can be viewed as
non-perturbative extensions of the usual Fluctuation-Dissipation Theorem (%
\textbf{FDT}) \cite{Kubo0,MPRV}. Such relations pertain either to
non-stationary transient situations \cite{Eva, Jarz} or to stationary
regimes \cite{Gal}. In particular, a family of fluctuation relations holds
for the distribution of work performed on a system \cite{Jarz, Crooks2, Sei,
Che1} which evolves in an equilibrium bath. This developments had an
important impact on the physics of nanosystems and biomolecules \cite{Ritort}%
. Here, we examine the question of the extension of such relations to an
evolution in a non-equilibrium medium. More precisely, we consider systems
placed in the thermal bath that is non-stationary, a situation which can be
realized experimentally by adding or extracting heat, or by modulating the
pressure. Let us remark that non-stationarity of the temperature is linked
via Stokes law to non-stationarity of friction and that particles with
time-dependent radii have non-constant mass and friction. This last
situation is an important problem in astrophysics for the formation of
planets through dust aggregation \cite{Blu1, Aus1}. There is another
situation where the non-stationarity of friction can be realized with
particles diffusing in ferroelectric fluids when external magnetic fields
are controlling the intrinsic viscosity \cite{Gun1}.

A famous example of a system which evolves in a non-stationary bath is the
Temperature Ratchet Model \cite{Rei1} of a Brownian motor. We can also
consider the evolution of a system initially in equilibrium at high
temperature (i.e. in the state with the probability density $\rho_{i}=\exp (-%
\frac{H}{T_{i}}))$ put in contact with a non stationary bath reaching a low
temperature $T_{f}$. The function $T(t)$ is then called the cooling
schedule. A particularly interesting case is the instantaneous quench of the
initial system in a thermostat at temperature $T_{f}$ which correspond to
the cooling schedule with instantaneous initial change of temperature. The
age (or waiting time) of the system is then the time elapsed since the
quench. Finally, at the more formal level, such non-stationary evolution
rules appear naturally from a stationary dynamics after system size or
coarse grained expansion \cite{Gar1}. Diffusion properties in a
non-stationary medium have been recently studied in the case of
unidimensional overdamped dynamics driven by multiplicative non stationary
noise \cite{Den1,Vit1}.

The present paper consists of five sections. Sect.\thinspace 2 sets the
stage and notations for the model of non-equilibrium and non-linear Langevin
dynamics that we consider. In Sect.\thinspace 3, we recall the main result
of \cite{Che1} and we list different time inversions that permit to obtain
fluctuation relations for entropy production. We observe that in this
context the work performed on the system no longer obeys the fluctuation
relations but that there still exists a functional that upon averaging gives
the free energy change in a non-equilibrium protocol. In Sect.\thinspace 4.
we explicitly construct fluctuation-dissipation relations that result from
the Taylor expansion of the last fluctuation relation. In particular, we
recover the generalized FDT \cite{Che3} around a non-equilibrium diffusion
and the usual FDT for the energy which results from a pulse of temperature 
\cite{Risk}. We discuss also the physical meaning of the effective
temperature of an harmonic oscillator in a non-stationary bath. Finally,
Sect.\thinspace 5 presents our conclusions.

\paragraph{Acknowlegements.}

The author thanks Krzysztof Gawedzki, François Delduc and Gregory Falkovich
for discussions and acknowledges support of the Koshland center for Basic
Research.

\section{Thermally Driven Non-Linear Langevin Dynamics}

\label{nel} The general dynamics that we consider is described by a
non-autonomous diffusion process that we call Langevin dynamics (or
non-linear Brownian motion) in a $d$-dimensional (phase-)space $E$. The
process obeys the stochastic differential equation (SDE) 
\begin{equation}
\dot{x}^{i}\ =\ -\Gamma _{t}^{ij}(x)\partial _{j}H_{t}(x)+\Pi
_{t}^{ij}(x)\partial _{j}H_{t}(x)\,+G_{t}^{i}(x)+w_{t}^{i}(x)+\,\eta
_{t}^{i}(x)  \label{sys}
\end{equation}
which should be interpreted in the Stratonovich convention. Here $H_{t}(x)$
is the Hamiltonian of the system (the time indexation corresponds to an
explicit time dependence), $\Gamma _{t}(x)$ is a family of non-negative
matrices, $\Pi _{t}(x)\,$\ is a family of antisymmetric matrices,\ $G_{t}(x)$
is an external force (or a shear) and $w_{t}(x)$ is an additional corrective
term which will be explicited in eq.\,(\ref{w}) below and which comes from
the $x$-dependence of $\Gamma _{t}$ and $\Pi _{t}$. Finally $\eta _{t}(x)$
is a white Gaussian vector field with mean zero and covariance 
\begin{equation}
\ \big\langle\eta _{t}^{i}(x)\eta _{s}^{j}(y)\big\rangle=2D_{t}^{ij}(x,y)%
\delta (t-s)\text{ \ \ with \ }D_{t}(x,x)\equiv d_{t}(x)=\frac{1}{\beta _{t}}%
\Gamma _{t}(x).\text{\ }
\end{equation}
The last equality, which contains the temperature of the bath $T_t=\frac{1}{%
\beta _{t}}$, is called the Einstein relation. In the case when the bath is
in equilibrium, this equation reflects the fact that the friction and the
noise are two dual effects of the interaction with the bath. In this non
stationary setup, the Einstein relation is valid under the assumption of
local thermal equilibrium, provided the thermalisation of the immediate
environment of the Brownian particle is much faster than the characteristic
variation of temperature.\ Let us underline that in this context the term
''non-linear'' concerns the non-homogeneous properties of $\Gamma _{t}(x)$
and $\Pi _{t}(x)$ (which may characterize non homogeneous properties of the
bath) and not the fact that equation\ (\ref{sys}) is non-linear. Such
non-linear properties appear naturally in many situations: non-ideal plasmas
and gases \cite{Kil1,Kil2}, ultra-cold clusters of atoms or molecules cooled
by interaction with laser radiation, active Brownian particles \cite{Ebe1}.\
Finally, the additional corrective\ term in (\ref{sys}) is given by the
expression 
\begin{equation}
w_{t}^{i}(x)=\partial _{y^{j}}D_{t}^{ij}(x,y)| _{x=y}-\,\frac{1}{\beta _{t}}%
\partial _{j}\Pi _{t}^{ij}(x).  \label{w}
\end{equation}
The presence of this term assures that in the case with stationary
Hamiltonian and temperature (i.e. $H_{t}=H$, $\beta _{t}=\beta )$ and
without external force (i.e $G=0),$ the Gibbs density $exp(-\beta H)$ is an
invariant density\footnote{%
In fact, the Gibbs density is then an equilibrium density: an invariant
density with vanishing modified probability current\ \cite{Che3}.}. The
presence of this term $w_{t}(x)$ in equation (\ref{sys}) can appear as a
makeshift arrangement, but it was extensively studied in the literature of
non linear Brownian motion \cite{Kil1,Kil2}. Note that $w_t$ vanishes in the
case of linear Brownian motion where $D _{t}(x,y)=D_{t}$ and $\Pi
_{t}(x)=\Pi_t$. We call the deterministic part of the second member of eq.\,(%
\ref{sys}): $\,$%
\begin{equation}
u_{t}(x)=-\Gamma _{t}^{ij}(x)\partial _{j}H_{t}(x)+\Pi _{t}^{ij}(x)\partial
_{j}H_{t}(x)\,+G_{t}^{i}(x)+w_{t}^{i}(x)  \label{d}
\end{equation}
the drift term. An elementary case of non-linear Brownian motion is the
Landau-Lifshitz-Bloch dynamics of a Brownian spin \cite{Garc} in an
effective magnetic field $B_{t}^{eff}(x)=-\nabla H_{t}$ (which can
incorporate interaction with other spins). It follows the dynamics 
\begin{equation}
\dot{x}=-x\times \nabla H_{t}+\lambda _{t}x\times \left( x\times \nabla
H_{t}\right) +G_{t}(x)+x\times \zeta _{t}\text{ \ with }\left\langle \zeta
_{t}^{i}\zeta _{s}^{j}\right\rangle =\frac{2\lambda _{t}}{\beta _{t}}\delta
^{ij}\delta (t-s).
\end{equation}
The first term on the right hand side is the precession term, the second one
is the damping term and the third one an external torque. The noise (and the
damping term) accounts for the effect of the interaction with the
microscopic degrees of freedom (phonons, conducting electrons, nuclear
spins...). This dynamics is a particular case of (\ref{sys}) with $\Gamma
_{t}^{ij}(x)=\lambda _{t}\left( \delta ^{ij}x^{2}-x^{i}x^{j}\ \right) $ and $%
\Pi _{t}^{ij}(x)=\varepsilon ^{ijk}x^{k}$ with $\varepsilon ^{ijk}$ the
totally antisymmetric tensor$.$ In this example one sees the need for the
term $\Pi _{t}^{ij}(x)\partial _{j}H_{t}(x)$ in eq.\,(\ref{sys})
corresponding to the Hamiltonian vector field which also permits to describe
systems of non-overdamped Brownian particles in the phase-space with
coordinates $\big( 
\begin{array}{c}
_q \\ 
^p
\end{array}
\big)$, non-stationary mass, in an external potential $V_{t}$, subjected to
a non-conservative force $f_{t}$ and in a non-stationary bath giving rise to
the noise and the non-homogeneous, non-linear but isotropic drag $\gamma
_{t}(q,p)$. The Stratonovich SDE which governs this model is then 
\begin{equation}
\left\{ 
\begin{array}{c}
\dot{q}=\frac{p}{m_{t}} \\ 
\text{ }\dot{p}=-\frac{\gamma _{t}(q,p)}{m_{t}}p-\nabla V_{t}(q)+f_{t}(q)+%
\frac{\nabla _{p}\gamma _{t}(q,p)}{2\beta _{t}}+\sqrt{\frac{2\gamma _{t}}{%
\beta _{t}}}\eta
\end{array}
\right. \text{\ \ \ with \ \ }\big\langle\eta _{t}^{i}\,\eta _{s}^{j}%
\big\rangle=\delta ^{ij}\delta (t-s)  \label{kr}
\end{equation}
which is again a sub-case of (\ref{sys}) with $\Gamma _{t}=\big( \, 
\begin{array}{cc}
0 & 0 \\ 
0 & \gamma _{t}
\end{array}
\big)$, $\Pi =\big( 
\begin{array}{cc}
0 & Id \\ 
-Id & 0
\end{array}
\big) ,$ $H_{t}\,=\,\frac{p^{2}}{^{2m_{t}}}+\,V_{t}(q)\,\ $and $G_{t}=%
\big( 
\begin{array}{c}
0 \\ 
f_{t}(q)
\end{array}
\big)$. The Kramers case corresponds to the Stokes law of friction $\gamma
_{t}(q,p)=\gamma _{t}(q)$. Another example of friction with $\gamma
_{t}(q,p)=\gamma (p^{2}-p_{0}^{2})$ appears in the Rayleigh-Helmholtz theory
of sound \cite{Ray}.

We start by collecting the elementary properties of diffusion processes that
we shall need \cite{Str}. \ The Markovian generator $\,L_{t}\,$ of the
process $\,x_{t}\,$ satisfying the SDE (\ref{sys}) is defined by the
relation 
\begin{equation}
\partial _{t}\big\langle f(x_{t})\big\rangle\,=\,\big\langle(L_{t}f)(x_{t})%
\big\rangle\text{ ,\ \ where \ \ }L_{t}\,=\,\widehat{u}_{t}\cdot \nabla
\,+\,\nabla \cdot \frac{\Gamma _{t}}{\beta t}\nabla,
\end{equation}
where the modified drift $\widehat{u}_t$ is defined in term of the drift (%
\ref{d}): 
\begin{equation}
\widehat{u}_{t}^{i}(x)=u_{t}^{i}(x)-\partial _{y^{j}} D_{t}^{ij}(x,y)|_{x=y}.
\label{uh}
\end{equation}
The time evolution of the instantaneous probability density function of the
process $\rho _{t}(x)=\left\langle \delta \left( x_{t}-x\right)
\right\rangle $ is governed by the formal adjoint $L_{t}^{\dagger }$ of the
generator $L_{t}$ 
\begin{equation}
\partial _{t}\rho _{t}=L_{t}^{\dagger }\rho _{t}  \label{FP}
\end{equation}
which can be rewritten as a continuity equation (resp. an hydrodynamic
advection equation) by defining the probability current $\widetilde{j_{t}}$
(resp. the mean local velocity $\widetilde{v_{t}}$:) 
\begin{equation}
\partial _{t}\rho _{t}=-\nabla\cdot\widetilde{j_{t}} =-\nabla \cdot\left(
\rho _{t}\widetilde{v_{t}}\right) \text{ \ \ where \ }\widetilde{j_{t}^{i}}%
\equiv ( \widehat{u}_{t}^{i}-\,\beta _{t}^{-1}\Gamma _{t}^{ij}\nabla _{j})
\rho _{t}\text{ \ and \ }\widetilde{v_{t}^{i}}\equiv \frac{\widetilde{%
j_{t}^{i}}}{\rho _{t}}.\text{ }  \label{hydro}
\end{equation}
As was explained in \cite{Che3}, it is convenient to use the freedom to add
a divergenceless term in the definition of the probability current to obtain
the modified current and the modified local velocity 
\begin{equation}
j_{t}^{i}\equiv \widetilde{j_{t}^{i}}+\beta _{t}^{-1}\nabla _{j}\left( \Pi
^{ij}\rho _{t}\right) \text{ \ \ and \ }v_{t}^{i}\equiv \frac{j_{t}^{i}}{%
\rho _{t}}  \label{v}
\end{equation}
which verify also the continuity equation (\ref{hydro}) but vanish in the
case of stationary Hamiltonian and temperature (i.e. $H_{t}=H$ , $\beta
_{t}=\beta )$ for vanishing external force $G_t=0$.

% With (\ref{d},\ref{uh},\ref{hydro},\ref
%{v}) we find the expression for the modified current:

%\begin{equation}
%j_{t}^{i}=\left( -\Gamma _{t}^{ij}\partial _{j}H_{t}+\Pi _{t}^{ij}\partial
%_{j}H_{t}\,+G_{t}^{i}-\beta _{t}^{-1}(\Gamma _{t}^{ij}-\Pi _{t}^{ij})\nabla
%_{j}\right) \rho _{t}.
%\end{equation}

\section{Fluctuation relations and time inversion}

In \cite{Che1}, various fluctuation relations were discussed for arbitrary
diffusion processes. We recall here the main result in the context of
systems with dynamics of type (\ref{sys}), see also Sect.\,3 of \cite{Fal2}.
With the use of combined Girsanov and Feynman-Kac formulae, one obtains the
detailed fluctuation relation (DFR) 
\begin{equation}
\mu _{\hspace{-0.01cm}_{0}}(dx)\,\,P_{\hspace{-0.05cm}_{T}}(x;dy,dW)\,\,%
\mathrm{e}^{-W}\,=\,\mu _{\hspace{-0.01cm}_{0}}^{r}(dy^{\ast })\,\,P_{%
\hspace{-0.05cm}_{T}}^{^{r}}(y^{\ast };dx^{\ast },d(-W))\,,  \label{DFR}
\end{equation}
where \hspace*{-0.3cm}{}

\begin{enumerate}
\item  $\,\mu _{\hspace{-0.01cm}_{0}}(dx)=\varrho _{\hspace{-0.01cm}%
_{0}}(x)\,dx\,$ is the initial distribution of the original forward process (%
\ref{sys}),

\item  $\,\mu _{\hspace{-0.01cm}_{0}}^{r}(dx)=\varrho _{\hspace{-0.01cm}%
_{0}}^{r}(x)\,dx\,$ is the initial distribution of the backward process
obtained from the forward process by applying a time inversion (see below),

\item  $\,P_{\hspace{-0.03cm}_{T}}(x;dy,dW)$ is the joint probability
distribution of the time $\,T\,$ position $\,x_{\hspace{-0.03cm}_{T}}\,$ of
the forward process starting at time zero at $\,x\,$ and of a functional $%
\,W_{\hspace{-0.03cm}_{T}}\,\ $(linked to the entropy production) of the
same process on the interval $\,[0,T]\,$ (described later),

\item  $\,P_{\hspace{-0.04cm}_{T}}^{^{r}}(x;dy,dW)$ is the similar joint
probability distribution for the backward process. %\end{itemize}
\end{enumerate}

\noindent The time inversion acts on time and space by an involution 
\begin{equation}
(t,x)\ \mapsto \ (t^{\ast }=T-t,x^{\ast })\,.  \label{stinv}
\end{equation}
Such an involution induces the action $\,\,x\mapsto \widetilde{x}\,\,$ on
trajectories by the formula $\,\widetilde{x}_{t}=x_{T-t}^{\ast }\,$ and,
further, the action on functionals of trajectories $\,\,F\mapsto \widetilde{F%
}\,$ \thinspace by setting $\,\widetilde{F}[x]=F[\widetilde{x}]$. \thinspace
To recover various fluctuation relations discussed in the literature \cite
{Kurchan,LebowSp,Crooks2,Jarz,SpS,CHCHJAR}, \thinspace one divides the drift
part (\ref{d}) of (\ref{sys}) into two parts, $\,u=u_{+}+u_{-}$, \thinspace
with $\,u_{+}\,$ transforming as a vector field under the space-time
involution (\ref{stinv}) and $\,u_{-}\,$ as a pseudo-vector field: 
\begin{equation}
u^{r}{}_{T-t,\pm }^{i}(x^{\ast })=\pm (\partial _{k}{x^{\ast }}%
^{i})(x)\,\,u_{t,\pm }^{k}(x)\,,\qquad u^{r}=u_{+}^{r}+\,u_{-}^{r}\,.
\label{inv}
\end{equation}
The random field $\,\eta _{t}(x)\,$ may be transformed with either rule. By
definition, the backward process satisfies then the Stratonovich SDE 
\begin{equation}
\dot{x}\ =\ u_{t}^{r}(x)\,+\,\eta _{t}^{r}(x)
\end{equation}
and, in general, differs from the naive time inversion $\,\widetilde{x}%
_{t}\, $ of the forward process. The functional $\,W_{\hspace{-0.02cm}%
_{T}}\, $ is given by the expression 
\begin{equation}
W_{\hspace{-0.02cm}_{T}}\ =\ln \varrho _{\hspace{-0.01cm}_{0}}(x_{\hspace{%
-0.02cm}_{0}})\ -\ln (\,\varrho _{0\hspace{-0.01cm}}^{r}(x_{T}^{\ast
})\sigma (x_{T}))\,+\,\int\limits_{0}^{T}\hspace{-0.1cm}J_{t}\,dt  \label{WT}
\end{equation}
where $\sigma (x)=\left| \det \left( \frac{\,\partial x^{\ast }}{\partial x}%
\right) \right| $ is the Jacobian of the spatial involution. \thinspace The
intensive functional $\,J_{t}\,$ has the interpretation of the rate of
entropy production in the environment and is given by the expression 
\begin{equation}
J_{t}\,=\,\beta _{t}\,\widehat{u}_{t,+}(x_{t})\cdot \Gamma _{t}^{-1}(x_{t})%
\big(\dot{x}_{t}-u_{t,-}(x_{t})\big)-(\nabla \cdot u_{t,-})(x_{t})\,,
\label{Jfat}
\end{equation}
\thinspace The time integral in eq.\,(\ref{WT}) is taken in the Stratonovich
sense.\thinspace When $\,\mu _{0\hspace{-0.01cm}}^{r}(dx^{\ast })=$ $\,\mu _{%
\hspace{-0.02cm}_{T}}(dx)$ then the boundary contribution $[\ln \varrho _{%
\hspace{-0.01cm}_{0}}(x_{\hspace{-0.02cm}_{0}})\ -\ln (\,\varrho _{0\hspace{%
-0.01cm}}^{r}(x_{T}^{\ast })\sigma (x_{T)})]$ to $\,W_{\hspace{-0.02cm}_{T}}$
gives the change in the instantaneous entropy of the process. In this case,
the functional $\,W_{\hspace{-0.02cm}_{T}}\,$ becomes equal to the overall
entropy production. Moreover, with the interpretation\ of $J_{t}$ as the
entropy production in the environment, the First Principle gives us that the
work $\mathcal{T}_{\hspace{-0.02cm}_{T}}$ performed on the system can be
expressed in term of $J_{t}$: 
\begin{equation}
\mathcal{T}_{\hspace{-0.02cm}_{T}}=H_{T}(x_{T})-H_{0}(x_{0})+\,\int%
\limits_{0}^{T}\hspace{-0.1cm}\frac{J_{t}\,}{\beta _{t}}dt.
\end{equation}
We can underline that in this setup, and contrary to the case of a the
stationary bath \cite{Che1}, the work $\mathcal{T}_{\hspace{-0.02cm}_{T}}$
cannot be identified with the functional $W_{\hspace{-0.02cm}_{T}}$ for an
appropriate choice of initial densities of the forward and backward
processes. This means that the work does not verify the DFR. \vskip 0.1cm

The DFR (\ref{DFR}) holds even if the measures $\,\mu_{\hspace{-0.01cm}%
_{0}}\,$ and $\,\mu _{\hspace{-0.01cm}_{0}}^{r}\,$ are not normalized, or
even not normalizable. \,When they are normalized, let us denote by $%
\big\langle-\big\rangle$ and by $\big\langle-\big\rangle^{r}$ the
expectations of functionals of, respectively, the forward and the backward
process on the time interval $[0,T]$, with initial distributions $\,\mu_{%
\hspace{-0.01cm}_{0}}\,$ and $\,\mu_{\hspace{-0.01cm}_{0}}^{r}\,$. 
%For normalized initial measures, we
%denote by 
%\begin{equation}
%\big\langle F\big\rangle\,=\,\int \mu _{\hspace{-0.01cm}%
%_{0}}(dx)E_{x}[F(x_{t})]\qquad \ \mathrm{and}\ \qquad \big\langle F%
%\big\rangle^{r}\,=\,\,\int \mu _{\hspace{-0.01cm}%
%_{0}}^{r}(dx)E_{x}^{r}[F(x_{t})]
%\end{equation}
%the averages over the realizations of the forward and the backward process $%
%\,x_{t},\ 0\leq t\leq T,\,$ with $\,x_{\hspace{-0.02cm}_{0}}\,$ distributed
%according to the probability measure $\,\mu _{\hspace{-0.01cm}_{0}}\,$ and $%
%\,\mu _{\hspace{-0.01cm}_{0}}^{r}$, \thinspace respectively. 
One of the immediate consequences of the DFR equation (\ref{DFR}) is the
(generalized) Jarzynski equality \cite{Jarz} 
\begin{equation}
\big\langle\mathrm{e}^{-W_{\hspace{-0.03cm}_{T}}}\big\rangle\ =\ 1
\label{JE}
\end{equation}
obtained by the integration of the both sides of eq.\,(\ref{DFR}). It
implies the inequality $\,\big\langle
W_{\hspace{-0.02cm}_{T}}\big\rangle\geq 0\,$ that has the form of the Second
Law of Thermodynamics stating the positivity of the average entropy
production. With a little more work \cite{Che1}, the DFR (\ref{DFR}) may be
cast into a form of the (generalized) Crooks relation \cite{Crooks2}: 
\begin{equation}
\big\langle F\exp(-W_{T})\big\rangle =\big\langle \widetilde{F}\big\rangle %
^{r}.  \label{Cro}
\end{equation}

We will now restrict ourselves to the class of time inversions (\ref{inv})
such that there exists a non-stationary density $f_{t}$ such that 
\begin{equation}
\widehat{u}_{+,t}=\beta _{t}^{-1}\Gamma _{t}\nabla \ln f_{t}\,\text{\ and
then \ }u_{-,t}\,=-\Gamma _{t}\nabla (H_{t}+\beta _{t}^{-1}\ln
f_{t})\,+\,\Pi _{t}\nabla H_{t}\,\,-\frac{1}{\beta _{t}}\nabla\cdot\Pi
_{t}^{T}+G_{t}\,.  \label{fam}
\end{equation}
After a straightforward calculation, the rate of entropy production in the
environment may be expressed as 
\begin{equation}
J_{t}=\dot{x}_{t}\cdot(\nabla \ln f_{t})(x_t)+((f_{t})^{-1}L_{t}^{\dagger }
f_{t})(x_t).
\end{equation}
With the choice $\mu _{0}(dx)=f_{0}(x)dx$ et $\mu _{0}^{r}(dx^{\ast
})=f_{T}(x)dx$ the functional $W_{T}$ takes then the simple form 
\begin{equation}
W_{T}=\int_{0}^{T}\left( (f_{t})^{-1}L_{t}^{\dagger }f_{t}-\partial _{t}\ln
(f_{t})\right)(x_t)\,dt.  \label{Wfat}
\end{equation}
We shall see that the fluctuation relations associated to this peculiar
family of inversions are the natural generalizations of the FDT. We shall
now describe particular cases in this family of time inversions.

\subsection{Complete reversal \protect\cite{Che1}}

As the function $f_{t}$ in (\ref{fam}) we take the instantaneous density
function (i.e. $f_{t}=\rho _{t}$) of the forward process (\ref{sys})
distributed with initial condition $f_{0}.$ Here, the functional (\ref{Wfat}%
) trivially vanishes\ $W_{T}=0$\ and the DFR (\ref{DFR}) takes the form of
the generalized detailed balance 
\begin{equation}
\mu _{\hspace{-0.01cm}_{0}}(dx)\,\,P_{\hspace{-0.05cm}_{T}}(x;dy)\,\,=\,\mu
_{\hspace{-0.01cm}_{T}}(dy)\,\,P_{\hspace{-0.05cm}_{T}}^{^{r}}(y^{\ast
};dx^{\ast })\,.
\end{equation}
One may show that $\rho _{t}^{r}(x)\equiv \rho _{t^{\ast }}(x^{\ast })$ is
the instantaneous density of the backward process and that the corresponding
probability current satisfies the relation 
\begin{equation}
\widetilde{j_{t}^{i,r}}(x)\,=\,-(\partial _{k}{x^{\ast }}^{i})(x)\,%
\widetilde{j^{k}}_{_{t^{\ast }}}(x^{\ast })\,.  \label{jjr}
\end{equation}
This inversion is employed in many articles in probability theory \cite
{Aze1,Mil1,Fol2, Pet1,Nel1}. It corresponds to the vanishing overall entropy
production.

\subsection{Current reversal \protect\cite{CHCHJAR,Che1}}

Another useful choice of time inversion, called the current reversal is
based on the choice $f_{t}=\pi _{t}\,$ where $\,\pi _{t}\,$ satisfies $\,\
L_{t}^{\dagger }\pi _{t}=-\nabla \cdot \widetilde{j_{t}}=0$. In the case
where $G_{t}=0$, we have $\pi _{t}=\exp (-\beta _{t}(H_{t}-F_{t}))$ with $%
F_{t}$ the free energy (i.e. $\exp (-\beta _{t}F_{t})=\int \exp (-\beta
_{t}H_{t})$ ). One can show \cite{Che1} that $\pi _{t}^{r}(x)\equiv \pi
_{t^{\ast }}(x^{\ast })$ is the density for the backward process which
correspond to the conserved current $\,\nabla \cdot \widetilde{j_{t}^{r}}=0$
and that eq.\,(\ref{jjr}) still holds. %we have 
%\begin{equation}
%\widetilde{j_{t}^{i,r}}(x)\,=\,-(\partial _{k}{x^{\ast }}^{i})(x)\,%
%\widetilde{j^{k}}_{_{t^{\ast }}}(x^{\ast })\,.
%\end{equation}
The functional \textbf{(}\ref{Wfat}\textbf{) }takes now the form 
\begin{equation}
W_{\hspace{-0.02cm}_{T}}^{ex}\,=\,-\,\int\limits_{0}^{T}(\partial _{t}\ln
\pi _{t})(x_{t})\,dt,  \label{WTC}
\end{equation}
where the index ''ex'' stands for ``excess'' \cite{Sei,Che1}. For the
backward process, the functional $W_{\hspace{-0.02cm}_{T}}^{ex,r}\,$is given
by the same expression with $\,\pi _{t}\,$ replaced by $\,\pi _{t}^{r}$.
\thinspace The Jarzynski equality (\ref{JE}) for this case was first proven
in one dimension in \cite{HatSas} and in the general case in \cite{Ge,
Liu1,Che1}.

\subsection{ \thinspace Canonical inversion \protect\cite{Che1}}

%%%%%%%%%%%%%%%%%%%%%%%%%%%%%%%%%%%%%%%%%%%%%%%%%%%%%%%%%%%%%%%%%%%%%%%%%%%%
%% Are you sure that the spurious terms are correctly taken into account %%%
%% below ?                                                               %%%
%%%%%%%%%%%%%%%%%%%%%%%%%%%%%%%%%%%%%%%%%%%%%%%%%%%%%%%%%%%%%%%%%%%%%%%%%%%%

A natural choice for the system (\ref{sys}) is to take $f_{t}$ to be the
Gibbs density $\exp (-\beta _{t}(H_{t}-F_{t}))$ where $F_{t}$ is the free
energy (i.e. $\exp (-\beta _{t}F_{t})=\int \exp (-\beta _{t}H_{t})$ ) if the
Gibbs density is normalizable and zero otherwise. This corresponds to the
choice $\widehat{u}_{+,t}=\,-\Gamma _{t}\nabla H_{t}$ in (\ref{fam}). The
functional \textbf{(}\ref{Wfat}\textbf{) }becomes 
\begin{equation}
W_{\hspace{-0.02cm}_{T}}^{ci}\,=-(\beta _{T}F_{T}-\beta
_{0}F_{0})+\,\int\limits_{0}^{T}\left[ \partial _{t}\left( \beta
_{t}H_{t}\right) +\beta _{t}G_{t}\cdot \nabla H_{t}-\nabla \cdot G_{t}\right]
(x_{t})\,dt\,  \label{Wci}
\end{equation}
where the index ''ci'' means ''canonical inversion''. The generalized
Jarzynski equality (\ref{JE}) can be rewritten in the form: 
\begin{equation}
\bigg\langle\exp \Big(-\int\limits_{0}^{T}\left[ \partial _{t}\left( \beta
_{t}H_{t}\right) +\beta _{t}G_{t}\cdot \nabla H_{t}-\nabla \cdot G_{t}\right]
(x_{t})\,dt\Big)\bigg\rangle=\exp \left[ -(\beta _{T}F_{T}-\beta _{0}F_{0})%
\right]   \label{efe}
\end{equation}
which permits to extract the difference of free energy out of a
non-equilibrium experiment in a non-stationary bath (but the connexion with
the work performed is lost). For example, for the Brownian particle (\ref{kr}%
), this functional takes the form: 
\begin{equation}
W_{\hspace{-0.02cm}_{T}}^{ci}\,=-(\beta _{T}F_{T}-\beta
_{0}F_{0})+\,\int\limits_{0}^{T}\Big[\dot{\beta _{t}}\Big(\frac{p_{t}^{2}}{%
^{2m_{t}}}+V_{t}(q_{t})\Big)+\beta _{t}\Big(-\frac{p_{t}\dot{m_{t}}}{%
m_{t}^{2}}+(\partial _{t}V_{t})(q_{t})+(f_{t}.\nabla V_{t})(q_{t})\Big)\Big]%
dt\,\,
\end{equation}
which, as compared to the functional which appears in the usual Jarzynski
equality \cite{Jarz, Kurchan}, contains new terms proportional to the
variation of temperature $\partial _{t}\beta _{t}$ and of mass $\partial
_{t}m_{t}$. From the Jarzynski equality (\ref{JE}), one can deduce the
inequality which constrains the evolution in a non-stationary bath 
\begin{equation*}
\left\langle W_{\hspace{-0.02cm}_{T}}^{ci}\right\rangle \geq 0\,.
\end{equation*}
In the case with a stationary Hamiltonian (i.e. $H_{t}=H$ ) and without
external force (i.e. $G_{t}=0)$ this constraint reads 
\begin{equation*}
\int_{0}^{T}(\partial _{t}\beta _{t})\big\langle H(x_{t})\big\rangle%
\,dt\,\geq \,(\beta _{T}F_{T}-\beta _{i}F_{i}).
\end{equation*}

Let us consider now\ the overdamped particle in a 3-dimensional
time-dependent harmonic potential. Such example admits an analytical
computation of the distribution of the functional $W_{\hspace{-0.02cm}%
_{T}}^{ci}.$ We consider a non stationary bath ($\gamma _{t},\beta _{t}$)
and the harmonic potential $U_{t}({x})=\frac{k_{t}}{2}({x}-{a}_{t})^{2}$
where $k_{t}$ is the stiffness coefficient and ${a}_{t}$ is the
instantaneous center of the potential. The particle is initially distributed
with the Gibbs density $\rho _{0}({x})=\exp (-\beta _{0}(U_{0}({x}%
)-F_{0}))=\left( \frac{C}{2\pi }\right) ^{\frac{3}{2}}\exp \left( -\frac{C}{2%
}({x}-{a}_{0})^{2}\right).$ Further, we will restrict our study to the
particular case, not necessarily physical, where the temperature of the bath
and the stiffness coefficient are such that their product is stationary: $%
k_{t}\beta _{t}=C.$ 
%%%%%%%%%%%%%%%%%%%%%%%%%%%%%%%%%%%%%%%%%%%%%%%%%%%%%%%%%%%%%%%%%%%%%%%%
%%%  Is it really a necessary restriction?                                       %%%%%
%%%%%%%%%%%%%%%%%%%%%%%%%%%%%%%%%%%%%%%%%%%%%%%%%%%%%%%%%%%%%%%%%%%%%%%%
This setup generalizes the unidimensional stationary case ($k_{t}=k$ , $%
\beta _{t}=\beta ,\gamma _{t}=\gamma $ and $a_{t}=ut$) considered in \cite
{Van1,Maz1}. The system satisfies the linear SDE 
\begin{equation}
\dot{x}=-\frac{k_{t}}{\gamma _{t}}\left( x-a_{t}\right) +\eta _{t}\text{ \ \
with \ \ }\big\langle \eta _{s}^{i}\eta _{t}^{j}\big\rangle =\frac{2}{\beta
_{t}\gamma _{t}}\delta (t-s)\delta ^{ij}.\text{ \ \ }  \label{exa}
\end{equation}
The functional $W_{T}^{ci}=$ $W_{\hspace{-0.02cm}_{T}}^{ex}$ takes here the
form 
\begin{equation*}
W_{T}^{ci}=-C\int_{0}^{T}\dot{a}_{t}\cdot\left( x_{t}-a_{t}\right) .
\end{equation*}
The distribution of $W_{T}^{ci}$ for a process with the initial Gaussian
density $\rho _{0}$ is Gaussian due to the linearity of\thinspace\ (\ref{exa}%
). A straightforward calculation gives the mean 
\begin{equation}
\left\langle W_{T}^{ci}\right\rangle =C\int_{0}^{T}dt\int_{0}^{t}\dot{a}%
_{t}\cdot\dot{a}_{s}\exp \left( -\int_{s}^{t}\frac{k_{u}}{\gamma _{u}}%
du\right)ds \text{ \ }  \label{exm}
\end{equation}
and the variance of this Gaussian (we assume that the integrals exist):

\begin{equation}
V_{T}\equiv \left\langle \left( W_{T}^{ci}-\left\langle
W_{T}^{ci}\right\rangle \right) ^{2}\right\rangle
=2C\int_{0}^{T}dt\int_{0}^{t}\dot{a}_{t}\cdot\dot{a}_{s}\exp \left(
-\int_{s}^{t}\frac{k_{u}}{\gamma _{u}}du\right)ds .\text{ \ }  \label{exv}
\end{equation}
The distribution of $W_{T}^{ci}$ is then 
\begin{equation}
P^{T}(W)=\frac{1}{\sqrt{2\pi V_{T}}}\exp \left( -\frac{\left( W-\left\langle
W_{T}^{ci}\right\rangle \right) ^{2}}{2V_{T}}\right)
\end{equation}
and an elementary calculus shows that the Jarzynski equality (\ref{JE}) is
equivalent to the fact that $V_{T}=2\left\langle W_{T}^{ci}\right\rangle $
which is evident from the comparison of (\ref{exm}) and (\ref{exv}).

\subsection{New inversion}

We choose for the function $f_{t\text{ }}$the mean instantaneous density $%
\rho _{t}^{\prime }$ of another Langevin dynamics (\ref{sys}) which
possesses the same parameters $\Gamma _{t}$, $\Pi _{t}$, $G_{t}$ but with
another non autonomous Hamiltonian $H_{t}^{\prime }$ and another bath
temperature $\beta _{t}^{^{\prime }}.$ We note $L_{t}$ (resp. $L_{t}^{\prime
}$ ) the Markovian generators\ of the process with the Hamiltonian $H_{t}$\
and the bath temperature $\beta _{t}$ (resp. $H_{t}^{\prime }$ and $\beta
_{t}^{^{\prime }}$). The functional \textbf{\ (}\ref{Wfat}\textbf{) } takes
now the form 
\begin{equation}
W_{T}=\int_{0}^{T}\big[(\rho _{t}^{\prime })^{-1}\left( L_{t}-L_{t}^{\prime
}\right) ^{\dagger }\rho _{t}^{\prime }\big](x_{t})\,dt.  \label{Wnew}
\end{equation}
This new inversion will permit to recover new generalizations of the FDT
around non-stationary non-equilibrium diffusions, see also \cite
{Che3,Fal2,Bae1}.

\section{ Generalizations of the Fluctuation-Dissipation Theorem}

%(FDT) obtained as a Taylor expansion of fluctuation relations}

As noted in \cite{Gall2,LebowSp}, the fluctuation relations may be viewed as
extensions to the non-perturbative regime of the Green-Kubo and Onsager
relations for the non-equilibrium transport coefficients valid within the
linear response description of the vicinity of the equilibrium.
Ref.\thinspace \cite{Che1} contains a detailed argument showing that if in a
stationary bath one perturbs an equilibrium system by introducing a weakly
time dependent Hamiltonian $H_{t}(x)=H(x)-g_{a,t}O^{a}(x)$ then the
Jarzynski equality associated to (\ref{WTC}) or (\ref{Wci}) gives in the
second order of the Taylor expansion in $g$ the usual FDT. Ref.\thinspace 
\cite{Fal2} showed that similar correspondence still holds around
non-equilibrium steady states for a stationary dynamics with an external
force (i.e. $G\neq 0$). In this case, it is the Crooks relation (\ref{Cro})
associated to the functional (\ref{WTC}) which gives the modified
Fluctuation-Dissipation Theorem \cite{Fal2,Gom1} after the first order
Taylor expansion in $g$. The second order expansion of Jarzynski equality (%
\ref{JE}) associated to the functional (\ref{WTC}) gives in such a situation
only a special case of this theorem. We shall now investigate which type of
fluctuation-dissipation identities may be deduced by Taylor expanding the
fluctuation relation corresponding to the time inversion of Sect 3.4.

\subsection{FDT around non-stationary diffusions}

We consider the system (\ref{sys}) with the Hamiltonian $%
H_{t}(x)=H_{t}^{0}(x)-g_{a,t}O^{a}(x)$. Following Sect.\thinspace 3.4, we
choose $f_{t}$ as the mean instantaneous density $\rho _{t}^{0}$ of the
unperturbed system with $g=0$. The functional (\ref{Wnew}) becomes 
\begin{equation}
W_{T}=\int_{0}^{T}g_{a,s}\left( (\rho _{s}^{0})^{-1}M_{s}^{a\dagger }\rho
_{s}^{0}\right) _{s}\,ds\text{ \ \ with }\ M_{s}^{a}=\left( \Gamma
_{s}\nabla O^{a}-\Pi _{s}\nabla O^{a}\right) \cdot \nabla ,\text{\ \ \ }
\end{equation}
where the subscript ``$s$'' on $\left( (\rho _{s}^{0})^{-1}M_{s}^{b\dagger
}\rho _{s}^{0}\right) $ signals that the latter function should be taken at
the point $x_{s}$, Let us now write a particular case of Crooks relation (%
\ref{Cro}), where the average is in the system (\ref{sys}) with the
Hamiltonian $H_{t}$, associated to a single time functional $%
F[x]=O^{a}(x_{t})\equiv O_{t}^{a}$ ($\,0<t<T$): 
\begin{equation}
\big\langle O_{t}^{a}\,\mathrm{e}^{-W_{T}}\big\rangle\,=\,\big\langle %
O_{T-t}^{a}\big\rangle^{r}\hspace{0.025cm}.\text{ \ }  \label{Crtop}
\end{equation}
We shall denote by $\left\langle \,.\,\right\rangle _{0}$ the average of the
process with the dynamics driven by $H_{t}^{0}$ and by $L^{0}$, $\widetilde{%
v^{0}}$ and $v^{0}$, respectively, its Markovian generator, its mean local
velocity and its modified mean local velocity. The first order Taylor
expansion 
\begin{equation}
\exp (-W_{T})=1+\int_{0}^{T}g_{b,s}\,\left( (\rho
_{s}^{0})^{-1}M_{s}^{b\dagger }\rho _{s}^{0}\right) _{s}\,ds\,+\,\mathcal{O}%
(g^{2})
\end{equation}
in (\ref{Crtop}) gives the relation 
\begin{equation}
\big\langle O_{t}^{a}\,\big\rangle_{0}+\int g_{b,s}\frac{\delta }{\delta
g_{b,s}}\Big|_{g=0}\,\left\langle O_{t}^{a}\right\rangle
\,ds\,-\int_{0}^{T}g_{b,s}\left\langle O_{t}^{a}\left( \rho
_{s}^{0}{}^{-1}M^{b\dagger }\rho _{s}^{0}\right) _{s}\right\rangle
_{0}ds\,+\,\mathcal{O}(g^{2})=\,\big\langle O_{T-t}\big\rangle^{r}.
\end{equation}
The right hand side has a functional dependence only on $\{g_{u},u>t\},$ so
if we apply $\,\frac{\delta }{\delta g_{b,s}}|_{g=0}\,$ for $\,0<s\leq t\,$
to the last identity, we obtain: 
\begin{equation}
\frac{\delta }{\delta g_{b,s}}\Big|_{g=0}\,\left\langle
O_{t}^{a}\right\rangle =\left\langle \left( (\rho _{s}^{0})^{-1}M^{b\dagger
}\rho _{s}^{0}\right) _{s}\,O_{t}^{a}\right\rangle _{0}.  \label{lin'}
\end{equation}
A short calculation gives: 
\begin{eqnarray}
(\rho _{s}^{0})^{-1}M_{s}^{b\dagger }\rho _{s}^{0} &=&\beta _{s}\widetilde{%
v_{s}^{0}}\cdot \nabla O^{b}-\beta _{s}L_{s}^{0}O_{s}^{b}+\Pi
_{s}^{ij}(\partial _{j}O^{b})\partial _{i}\ln (\rho _{s}^{0})+(\partial
_{i}\Pi _{s}^{ij})\partial _{j}O^{b}  \label{dec1} \\
&=&\beta _{s}\left( 2\widetilde{v_{s}^{0}}\cdot \nabla -L_{s}^{0}\right)
O^{b}-\beta _{s}\widetilde{v_{s}^{0}}\cdot \nabla O^{b}+\Pi
_{s}^{ij}(\partial _{j}O^{b})\partial _{i}\ln (\rho _{s})+(\partial _{i}\Pi
_{s}^{ij})\partial _{j}O^{b}  \notag \\
&=&\beta _{s}\left( 2\widetilde{v_{s}^{0}}\cdot \nabla -L_{s}^{0}\right)
O^{b}-\beta _{s}v_{s}^{0}\cdot \nabla O^{b}.  \notag
\end{eqnarray}
Moreover, we have the sum rule (for $s\leq t)$ 
\begin{equation}
\partial _{s}\left\langle O_{s}^{b}\,O_{t}^{a}\right\rangle
_{0}=\left\langle \big((2\widetilde{v_{s}^{0}}\nabla -L_{s}^{0})O^{b}\big)%
_{s}\,O_{t}^{a}\right\rangle _{0}.  \label{regsom}
\end{equation}
With (\ref{lin'}), (\ref{regsom}) and(\ref{dec1}), we obtain the Modified
Fluctuation-Dissipation Theorem : 
\begin{equation}
\partial _{s}\left\langle O_{s}^{b}\,O_{t}^{a}\right\rangle _{0}=\frac{1}{%
\beta _{s}}\frac{\delta }{\delta g_{b,s}}\Big|_{g=0}\,\left\langle
O_{t}^{a}\right\rangle +\left\langle (v_{s}^{0}\cdot \nabla
O^{b})_{s}\,O_{t}^{a}\right\rangle _{0}.  \label{MFDT}
\end{equation}
This is a generalization of the FDT around a non-equilibrium diffusion
process in a stationary bath ($\beta _{t}=\beta )$ of refs. \cite
{Fal2,Che3,Bae1,Cug1} and of FDT around non-stationary Langevin equation of
ref. \cite{Zim1}. It may be also proven as in \cite{Che3} using the fact
that the diffusion process becomes an equilibrium one in the Lagrangian
frame of its modified mean local velocity $v^{0}$, verifying in that frame
the usual FDT. The transformation of the latter back to the Eulerian (i.e.
laboratory) frame leads to (\ref{MFDT}). Let us remark that here, similarly
as in the stationary case discussed in \cite{Fal2}, the Jarzynski equality (%
\ref{JE}) for the functional (\ref{Wnew}) leads upon the second order
expansion in $g$ to a particular case of the \textbf{MFDT} where the
observable $O_{t}^{a}$ is replaced by 
% but to particular case %of this. This can be shown 
%by developing $\exp (-W_{T})$ in second order in $%
%g$: 
%\begin{equation}
%\exp (-W_{T})=1-\int_{0}^{T}dsg_{a,s}{}\rho _{s}^{0}{}^{-1}M_{s}^{a\dagger
%}\rho _{s}^{0}+\frac{1}{2}\int_{0}^{T}\int_{0}^{T}dsdtg_{a,s}g_{b,t}\left(
%\rho _{s}^{0}{}^{-1}M_{s}^{a\dagger }\rho _{s}^{0}\right) \left( \rho
%_{t}^{0}{}^{-1}M_{t}^{b\dagger }\rho _{t}^{0}\right) +\mathcal{O}(g^{3}).
%\end{equation}
%Then using the same trick that before we obtain 
%\begin{equation}
%\frac{\delta }{\delta g_{b,s}}\Big|_{g=0}\,\left\langle (\rho
%_{t}^{0})^{-1}M_{t}^{a\dagger }\rho _{t}^{0}\right\rangle =\left\langle
%\left( (\rho _{s}^{0})^{-1}M^{a\dagger }\rho _{s}^{0}\right) \left( (\rho
%_{t}^{0})^{-1}M_{t}^{a\dagger }\rho _{t}^{0}\right) \right\rangle _{0}
%\end{equation}
%which can be written and the form of the MFDT 
%\begin{equation}
%\frac{1}{\beta _{s}}\frac{\delta }{\delta g_{b,s}}\Big|_{g=0}\,\left\langle
%A_{t}^{a}\right\rangle =\partial _{s}\left\langle
%O_{s}^{b}A_{t}^{a}\right\rangle _{0}-\left\langle v_{s}^{0}\nabla
%O_{s}^{b}A_{t}^{a}\right\rangle _{0}
%\end{equation}
%but for the observable $A^a_t$ where 
$A^{a}=(\rho _{t}^{0})^{-1}M_{t}^{a\dagger }\rho _{t}^{0}$ which is a
(time-dependent) functional of $O^{a}$. \vskip0.1cm

The violation of the usual FDT can be parametrized by using (\ref{MFDT}) via
the introduction of the so-called effective temperature \cite{Cug2,Cri1}
defined by 
\begin{equation}
T^{eff}(s,t,O^{a})\equiv \frac{\partial _{s}\left\langle
O_{s}^{a}O_{t}^{a}\right\rangle _{0}}{\frac{\delta }{\delta g_{a,s}}\Big|%
_{g=0}\,\left\langle O_{t}^{a}\right\rangle }=\frac{1}{\beta _{s}}+\frac{%
\left\langle (v_{s}^{0}\cdot \nabla O^{a})_{s}\,O_{t}^{a}\right\rangle _{0}}{%
\frac{\delta }{\delta g_{a,s}}\Big|_{g=0}\,\left\langle
O_{t}^{a}\right\rangle }.  \label{Teff}
\end{equation}
We shall consider now the case where this effective temperature may be
computed analytically in order to verify its physical consistency.

\paragraph{Unidimensional harmonic oscillator in a non-stationary bath.}

The SDE which governs this systems is

\begin{equation}
\dot{x}=-\frac{k}{\gamma }x+\,\eta \,,\text{ \ \ \ with \ \ \ }\big\langle%
\eta _{t}\,\eta _{s}\big\rangle=\frac{2T_{t}}{\gamma }\delta (t-s).
\label{ho}
\end{equation}
We take the cooling schedule $T_{t}$ such that the bath passes from an
initial temperature ($T_{0}=T_{i}$) to a lower final temperature ($%
T_{\tau}=T_{f}<T_{i}$) during a time $\tau.$ The system is initially in
equilibrium with the bath and its initial density is $\rho _{0}(x)= \exp (-%
\frac{k}{2T_{i}}x^{2})/Z_{i}.$ We consider two particular examples of
cooling schedules:

\begin{itemize}
\item  Instantaneous quench 
\begin{equation}
T_{t}=\left\{ 
\begin{array}{c}
T_{i}\text{\ \ \ \thinspace \thinspace if \ }t=0 \\ 
T_{f}\text{ \ \ \ if \ }t>0
\end{array}
\right. .
\end{equation}

\item  Linear decrease of temperature: \ 
\begin{equation}
T_{t}=\left\{ 
\begin{array}{c}
T_{i}+\frac{t}{\tau }(T_{f}-T_{i})\text{ \ \ \ \ \ \thinspace \thinspace if
\ \ }t\leq \tau \\ 
T_{f}\text{ \ \ \ \ \ \ \ \ \ \ \ \ \ \ \ \ \ \ \ \ \ \ \ if\ \ \ }t\geq \tau
\end{array}
\right. .
\end{equation}
\end{itemize}

Due to the linearity of eq.\,(\ref{ho}), one can compute explicitly the
response of the position to an external perturbation $V(x)=\frac{k}{2}%
x^{2}\rightarrow $ $V_{t}^{\prime }(x)=V(x)-g_{t}x$ with $g_{0}=0$: 
\begin{equation}
\frac{\delta }{\delta g_{s}}\Big|_{g=0}\,\left\langle x_{t}\right\rangle =%
\frac{1}{\gamma }\exp\left(-\frac{k}{\gamma }(t-s)\right).  \label{rspf}
\end{equation}
It has a stationary form and is independent on the cooling schedule. In a
similar way, we obtain also an explicit expression for the dynamical 2-time
correlation function of the position in the unperturbed system in the two
cooling schedules. For $s\leq t$, we obtain for the instantaneous quench: 
\begin{equation}
\left\langle x_{s}x_{t}\right\rangle _{0}=\frac{T_i-T_f}{k} \exp \left( -%
\frac{k}{\gamma }(s+t)\right) +\frac{T_{f}}{k}\exp \left( -\frac{k}{\gamma }%
(t-s)\right)  \label{co}
\end{equation}
and for the linear decrease of temperature schedule: 
\begin{equation}
\left\langle x_{s}x_{t}\right\rangle _{0}=\left\{ 
\begin{array}{c}
\left( \frac{T_{i}}{k}+\frac{T_{f}-T_{i}}{k\tau }\left( s- \frac{\gamma}{2k}%
\right) \right) \exp \left( -\frac{k}{\gamma }(t-s)\right) +\frac{\gamma
(T_{f}-T_{i})}{2k^{2}\tau }\exp (-\frac{k}{\gamma }(s+t))\ \text{ if\ }s\leq
\tau,\ s\leq t \\ 
\frac{T_{f}}{k}\exp \left( -\frac{k}{\gamma }(t-s)\right) -\frac{%
\gamma(T_{f}-T_{i}) }{2k^{2}\tau }\left( \exp (\frac{2k}{\gamma }\tau
)-1\right) \exp \left( -\frac{k}{\gamma }(s+t)\right)\ \,\text{\ if }\tau
\leq s\leq t.
\end{array}
\right.  \label{cof}
\end{equation}
We see in these two formulae that the characteristic time of convergence
toward the Gibbs density $\exp (-\frac{k}{2T_{f}}x^{2})/Z_{f}$ is $\frac{%
\gamma }{k}$ for the\ instantaneous quench and $\tau +\frac{\gamma }{k} $
for the linear decrease schedule. Note the relation 
\begin{equation}
\left\langle x_{s}x_{t}\right\rangle _{0}\,=\, \left\langle
x_{s}^2\right\rangle _{0}\,\exp\left(-\frac{k}{\gamma}(t-s)\right)
\,=\,\left\langle x_{s}^2\right\rangle _{0}\gamma\, \frac{\delta }{\delta
g_{s}}\Big|_{g=0}\left\langle x_{t}\right\rangle  \label{rapp}
\end{equation}
holding for both cooling schedules. It shows that at very large times (i.e. $%
t>s\gg \tau ,\frac{\gamma }{k}$ ) the correlation functions (\ref{co}) and (%
\ref{cof}) take a stationary form depending on $t-s$. The instantaneous mean
density of the process is Gaussian at all times. It follows that the mean
local velocity has the form 
\begin{equation}
v^0_s(x)\,=\,\Big(-\frac{k}{\gamma}+\frac{T_s}{\gamma\left\langle
x_{s}^2\right\rangle _{0}}\Big)\,x
\end{equation}
and the corrective term in the {FDT} (\ref{MFDT}) is 
\begin{equation}
\left\langle (v_s^0\cdot\nabla x)_s\,x_{t}\right\rangle _{0}\,=\,
\left\langle (v_s^0)_s\,x_{t}\right\rangle _{0}\,=\, \Big(-\frac{k}{\gamma}+%
\frac{T_s}{\gamma\left\langle x_{s}^2\right \rangle _{0}}\Big)\left\langle
x_{s}x_t\right\rangle _{0}  \label{corrt}
\end{equation}
Using the relations (\ref{rapp}) and (\ref{corrt}), the {FDT} (\ref{MFDT})
may be rewritten in this case as the identity 
\begin{equation}
\Big(\partial_s+\frac{k}{\gamma}\Big)\left\langle x_{s}x_t\right\rangle _{0}
\,=\,2T_s\frac{\delta }{\delta g_{s}}\Big|_{g=0} \left\langle
x_t\right\rangle
\end{equation}
which is easy to check directly. \vskip 0.1cm

The effective temperatures for the instantaneous quench $T_{Q}^{eff}$ and
for the linear decrease of temperature schedule $T_{LD}^{eff}$ are: 
\begin{equation}
T_{Q}^{eff}(s,t,x)=T_{f}+\left( T_{f}-T_{i}\right) \exp (-\frac{2k}{\gamma }%
s)\text{ \ \ \ if\ \ }0<s\leq t
\end{equation}
and 
\begin{equation}
T_{LD}^{eff}(s,t,x)=\left\{ 
\begin{array}{c}
T_{i}+\frac{1}{\tau }\left( T_{f}-T_{i}\right) \left( s+\frac{\gamma }{2k}%
\left( 1-\exp (-\frac{2k}{\gamma }s)\right) \right)\ \ \ \,\text{ if\ }\
0<s\leq \tau, \ s\leq t \\ 
T_{f}+\left( T_{f}-T_{i}\right) \exp \left(-\frac{2k}{\gamma }s\right) \frac{%
\exp (\frac{2k}{\gamma }\tau )-1}{\frac{2k}{\gamma }\tau }\text{ \ \ \ \,if
\ }\tau \leq s\leq t.
\end{array}
\right.  \label{Tefflc}
\end{equation}
The two effective temperatures have an expected behavior for large time $s$
converging toward $T_{f}$. However, we may see in this system the problems
with the physical interpretation of the effective temperature \cite
{Cug2,Cri1}. For example, $\lim_{s\rightarrow 0}T_{Q}^{eff}=2T_{f}-$ $T_{i}$ 
$\neq T_{i\text{ }}$ and this expression can be negative if $T_{f}<\frac{%
T_{i}}{2}.$ The possibility to find negative effective temperature has been
observed also in \cite{Fal2} and, for the kinetically constrained model, in 
\cite{May1}. Moreover, the effective temperature grows toward its limit $%
T_{f}$, which does not correspond to the physical intuition for the
temperature of a cooled system. The investigation of the linear-decrease
cooling schedule is instructive for the understanding of these two problems.
For this schedule, $\lim_{s\rightarrow 0}T_{LD}^{eff}=T_{i\text{ }}$ and the
effective temperature decreases from this value until time $\tau $ 
%$\left( \partial
%_{s}T_{LD}^{eff}(s<\tau )=\frac{1}{\tau }\left( T_{f}-T_{i}\right) \left(
%1+\exp (-\frac{2k}{\gamma }s)\right) <0\right) $ 
when it reaches $T_{LD}^{eff}(\tau )=$ $T_{f}+\left( T_{f}-T_{i}\right) 
\frac{1-\exp (-\frac{2k}{\gamma }\tau )}{\frac{2k}{\gamma }\tau }< T_f$.
\,So, the problem with the initial time limit of $T_{Q}^{eff}$ was due to
the instantaneous modeling of the quench. On the other hand, the second part
of the evolution for the linear cooling schedule (i.e.\,for $s\geq\tau)$
begins with an effective temperature below $T_{f}$ that may be even
negative. The last features are not really physically satisfying but the
first one explains why the effective temperature $T^{eff}_Q$ converges
toward $T_{f}$ by growing, the fact which is confirmed by (\ref{Tefflc}). We
represent below in Figure 1 typical joint evolution for the linear cooling
schedule of the bath temperature $T$ (the crosses) and the effective
temperature $T_{LD}^{eff}$ (the solid lines) in the case where $T_{i}=300K$, 
$T_{f}=200K$ and $\frac{2k}{\gamma }=1 s^{-1}$ for $\tau =0.1 s$ (red) $\tau
=1 s$(blue) and $\tau =10 s$ (black). The singularity of the limit $\tau
\rightarrow 0$ is evident on this graph.

\begin{figure}[ht]
\begin{center}
\includegraphics[scale=0.5]{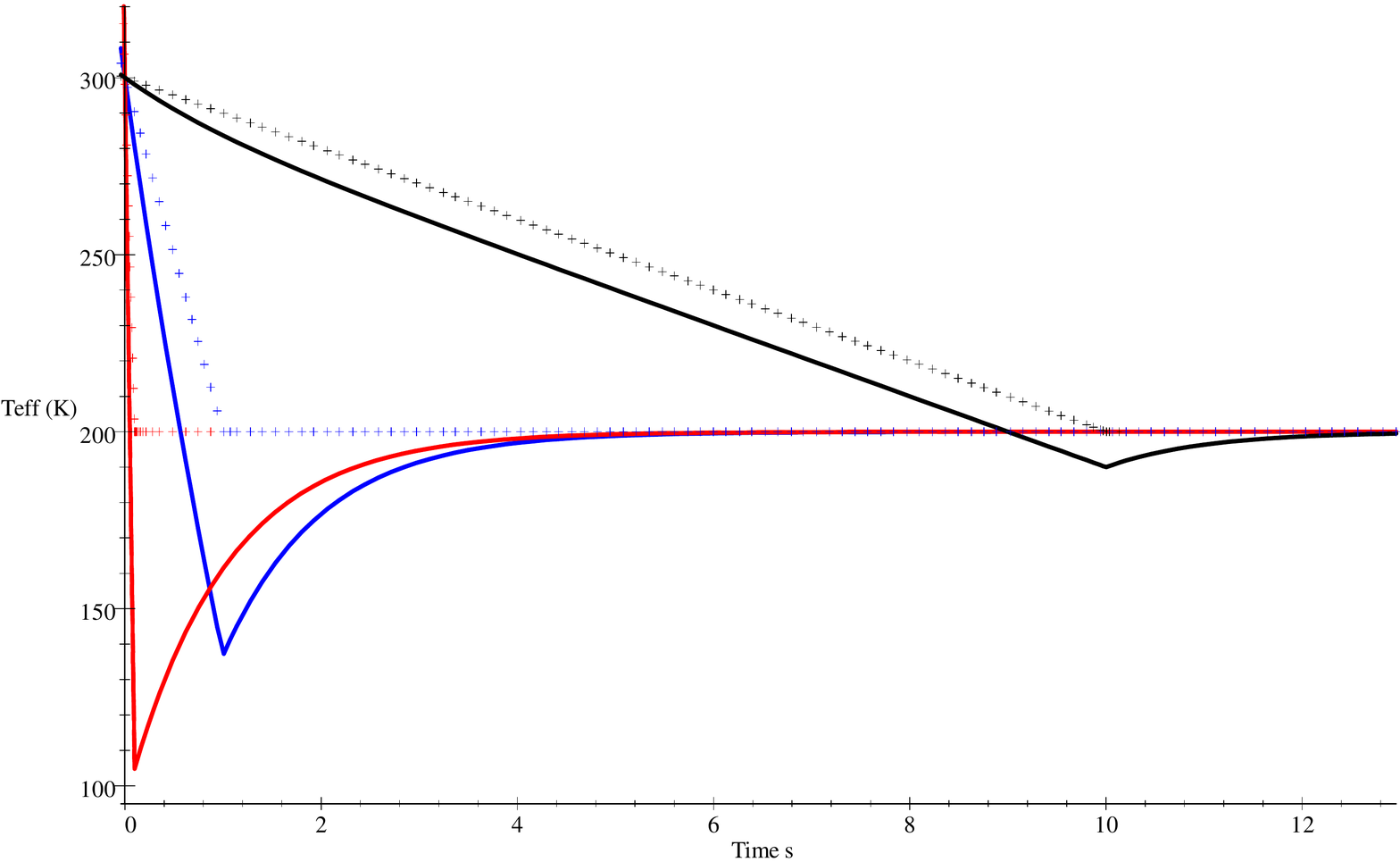}
\end{center}
\caption{ Red solid: $T_{LD}^{eff}$ for $\protect\tau =0.1 s$. Red crosses: $%
T$ for $\protect\tau =0.1 s$. Blue solid: $T_{LD}^{eff}$ for $\protect\tau %
=1 s$. Blue crosses: $T$ for $\protect\tau =1 s$. Black solid: $T_{LD}^{eff}$
for $\protect\tau =10 s$. Black crosses: $T$ for $\protect\tau =10 s$.}
\end{figure}

\subsection{Response of a diffusion to a pulse of bath temperature}

Let us consider the system whose dynamics is governed by eq.\thinspace (\ref
{sys}) with the variable bath temperature $\,\beta _{t}^{-1}=(1+g_{t})\beta
_{0}^{-1}$. We choose as the function $f_{t}$ the mean instantaneous density 
$\rho _{t}^{0}$ of the similar system with the constant bath temperature $%
\beta _{0}^{-1}$. The functional (\ref{Wnew}) becomes 
\begin{equation}
W_{T}=\int_{0}^{T}g_{s}\big((\rho _{s}^{0})^{-1}M_{s}^{\dagger }\rho _{s}^{0}%
\big)_{s}\,ds\text{ \ \ with }\ M_{s}=\left( \beta _{0}\right) ^{-1}\left(
\left( -\nabla _{j}\Pi _{s}^{ij}+\nabla _{j}\Gamma _{s}^{ij}\right) \nabla
_{i}+\Gamma _{s}^{ij}\text{\ }\nabla _{i}\nabla _{j}\right) .\text{\ \ }
\end{equation}
With the same reasoning as in Sect.\thinspace 4.1, we find the link between
the response to a pulse of temperature at time $s$ and the dynamical
correlation function in the system with stationary inverse temperature $%
\beta _{0}$: 
\begin{equation}
\frac{\delta }{\delta g_{s}}\Big|_{g=0}\,\left\langle O_{t}\right\rangle
=\left\langle \left( (\rho _{s}^{0})^{-1}M^{\dagger }\rho _{s}^{0}\right)
_{s}\,O_{t}(x_{t})\right\rangle _{0}.
\end{equation}
Here, there does not seem to exist a simplification of this relation in the
spirit of (\ref{MFDT}) and we cannot say more except for the case when the
system with the bath temperature $\beta _{0}$ is an equilibrium one.

\paragraph{Temperature pulse around equilibrium.\ }

In the case where the system with the bath temperature $\beta_0$ is in
equilibrium (i.e. without external force $G_{t}=0$ and with a stationary
Hamiltonian $H_{t}=H$ and the Gibbsian instantaneous density),\ the
functional (\ref{Wci}) takes the form: 
\begin{equation}
W_{\hspace{-0.02cm}_{T}}^{ci}\,=W_{T}^{ex}=-(\beta _{T}F_{T}-\beta
_{0}F_{0})+\,\int\limits_{0}^{T}\dot{\beta _{t}}H(x_{t})\,dt.\,\,
\end{equation}
We want to prove that the Taylor expansion in the second order of the
Jarzynski equality (\ref{JE}) associated to this functional gives the usual
Fluctuation-Dissipation Theorem for the energy \cite{Risk} 
\begin{equation}
\partial _{s}\left\langle H_{s}H_{t}\right\rangle _{0}\,=\,\frac{1}{\beta
_{0}}\frac{\delta }{\delta g_{s}}\Big|_{g=0}\,\left\langle
H_{t}\right\rangle.  \label{fdte}
\end{equation}
The equality (\ref{JE}) takes now the form 
\begin{equation}
\left\langle \exp \Big( -\int_{0}^{T}\dot{\beta _{t}}H(x_{t})\,dt\Big) %
\right\rangle =\frac{\int\exp \left( -\beta _{T}H(x)\right)dx }{\int\exp
\left(-\beta _{0}H(x)\right)dx}.  \label{oste}
\end{equation}
We develop the left member in second order in $g_t$ or $h_t=g_t-g_t^2$
assuming that $g_t$ vanishes for $t\leq0$: 
\begin{eqnarray*}
&&\left\langle \exp \Big( -\int_{0}^{T}\dot{\beta _{t}}H(x_{t})\,dt\Big) %
\right\rangle =\left\langle \exp \Big( \beta _{0}\int_{0}^{T} \dot{h_{t}}
H(x_{t})\,dt\Big)+\mathcal{O}(h^{3}) \right\rangle \\
&&=\left\langle 1+\beta _{0}\int_{0}^{T}\dot{h_{t}}H(x_{t})\,dt+\frac{\beta
_{0}^{2}}{2}\int_{0}^{T}dt\int_{0}^{T}\dot{h_{t}}\dot{h_{s}}%
H(x_{t})H(x_{s})\,ds+\mathcal{O}(h^{3})\right\rangle \\
&&=1+\beta _{0}\int_{0}^{T}\dot{h_{t}}\left\langle H(x_{t})\right\rangle
_{0}dt +\beta _{0}\int_{0}^{T}dt\int_{0}^{t}\dot{h_{t}}h_{u}\frac{\delta }{%
\delta g_{u}}\Big|_{g=0}\,\left\langle H_{t}\right\rangle\,du \\
&&+\frac{\beta _{0}^{2}}{2}\int_{0}^{T}dt\int_{0}^{T}\dot{h_{t}}\dot{h_{s}}%
\left\langle H(x_{t})H(x_{s})\right\rangle _{0}ds+\mathcal{O}(h^{3}) \\
&&=1+\beta _{0}\int_{0}^{T}\dot{h_{t}}\left\langle H(x_{t})\right\rangle
_{0}dt +\beta _{0}\int_{0}^{T}dt\int_0^T\dot{h_{t}}\dot{h_{s}}%
\theta(t-s)\,ds \int_s^t\frac{\delta }{\delta g_{u}}\Big|_{g=0}\,\left%
\langle H_{t}\right\rangle\,du \\
&&+\frac{\beta _{0}^{2}}{2}\int_{0}^{T}dt\int_{0}^{T}\dot{h_{t}}\dot{h_{s}}%
\left\langle H(x_{t})H(x_{s})\right\rangle _{0}ds+\mathcal{O}(h^{3}),
\end{eqnarray*}
where the last equality was obtained by expressing $h_u=\int_0^u\dot{h_s}ds$
in the second term and changing the order of integration over $s$ and $u$.
Expansion of the right member of eq.\,(\ref{oste}) gives in turn 
\begin{eqnarray*}
&&\frac{\int\exp \left( -\beta _{T} H(x)\right)dx }{\int\exp \left( -\beta
_{0}H(x)\right)dx }=1+\beta _{0}h_{T}\left\langle H\right\rangle_0+\frac{1}{2%
}(\beta _{0}h_{T})^2 \left\langle H^2\right\rangle_0 +\mathcal{O}(h^{3}) \\
&&=1+\beta_0\int_0^T\dot{h_t}\left\langle H(x_t)\right\rangle_0dt+ \frac{%
\beta _{0}^2}{2}\int_0^Tdt\int_0^T\dot{h_t}\dot{h_s} \left\langle
H(x_t)^2\right\rangle_0ds.
\end{eqnarray*}
The comparison of the terms quadratic in $\dot h$ leads to the identity 
\begin{equation}
\frac{1}{\beta _{0}}\int_{s}^{t}du\frac{\delta }{\delta g_{u}}\Big|%
_{g=0}\,\left\langle H_{t}\right\rangle = \left\langle H_{t}^2\right\rangle
_{0}-\,\left\langle H_{s}H_{t}\right\rangle _{0}
\end{equation}
for $s\leq t$ which gives the relation (\ref{fdte}) by the derivation with
respect to $s$. Once again the Jarzynski equality appears as a global
version of the FDT.

\section{Conclusions}

\bigskip We have discussed fluctuation relations for diffusion processes (%
\ref{sys}) in a non-stationary thermal bath. Those included the fluctuation
relations for the entropy production (\ref{DFR}). The work performed on the
system no longer verifies such fluctuation relations, but that there still
exists a relation (\ref{efe}) which permits to extract the free energy
difference in a non-equilibrium experiment. We proved that the fluctuation
relations involving the functional (\ref{Wnew}) are global versions of the
Modified Fluctuation-Dissipation Theorem (\textbf{MFDT}) (\ref{MFDT}) around
a non-equilibrium diffusion extending the \textbf{MFDT} obtained before in 
\cite{Fal2,Che3} and of the usual \textbf{FDT} for energy (\ref{fdte}) \cite
{Risk} resulting from a pulse of temperature. On the way, in Sec 4.1 we
illustrated the extended \textbf{MFDT} on a simple example of a harmonic
oscillator in a thermal bath with variable temperature and we investigated
the physical meaning of the effective temperature introduced in \cite
{Cug2,Cri1} for such a system. One should underline that the interaction
with a non-stationary bath is one among many ways to thermally drive a
system. For example the thermodiffusion effect (or Sorret effect) which
appears in a bath with non-uniform temperature (i.e $\beta (x)=\frac{1}{T(x)}
$) has been explained in \cite{Van}, for the unidimensional case, using a
stationary non-equilibrium microscopic model of the type (\ref{sys}) with $%
G=0$ but with the thermophoretic force $-\Gamma \frac{dT}{dx}$ added to the
drift (\ref{d}). In the same spirit, many years ago, Landauer \cite{Lan}
proposed a model with the wall temperature varying along a very narrow pipe
filled with the Knudsen gas described by a stationary non-equilibrium
microscopic model of the type (\ref{sys}) but with the thermophoretic force $%
-\Gamma \frac{dT}{dx}$ and the chemical force $-T\frac{d\Gamma }{dx}$ added
to the drift. Finally, other way to drive a system is to consider
fluctuating coefficients in (\ref{sys}), for example \cite{Aus1} considered
a fluctuating mass and \cite{Roz1,Luc1,Tal1} a stochastic friction. It would
be interesting to describe the fluctuation relations in those setups.

\end{document}